\documentclass[mathleft
]{an}
\usepackage{graphicx}
\usepackage{times}
\usepackage{natbib}

\usepackage{amsmath,amssymb}
\graphicspath{{./figs/}}

\newcommand\cyg{\mbox{Cygnus~X-1}}

\newcommand\gaphi{\gamma^{(\phi)}}
\newcommand\pli{\ensuremath{\Gamma}}
\newcommand\de{\ensuremath{\delta}}
\newcommand\gi{\ensuremath{g_\mathrm{lp}}}
\newcommand\dr{\ensuremath{\Delta r}}
\newcommand\ri{\ensuremath{r}}
\newcommand\rg{\ensuremath{r_\mathrm{g}}} 

\begin{document}

\Pagespan{0}{}
\Yearpublication{2016}%
\Yearsubmission{2016}%
\Month{01}%
\Volume{000}%
\Issue{00}%
\DOI{This.is/not.aDOI}%

\title{Relativistic Reflection: Review and Recent Developments in Modeling}

\author{T. Dauser\inst{1}\fnmsep\thanks{Corresponding author:
  \email{thomas.dauser@fau.de}\newline}
  \and  J.~Garc\'ia \inst{2}
  \and J.~Wilms \inst{1}
}
\titlerunning{Review on Relativstic Reflection}
\authorrunning{T.~Dauser, J.~Garc\'ia, and J.~Wilms}
\institute{
  Dr.\ Karl Remeis-Observatory and Erlangen Centre for
  Astroparticle Physics,
  Sternwartstr.~7, 96049 Bamberg, Germany 
\and 
  Harvard-Smithsonian Center for Astrophysics, 60 Garden
  Street, Cambridge, MA 02138, USA
}

\received{01 Jan 2016}
\accepted{15 Jan 2016}
\publonline{2016 Apr 25}

\keywords{black hole physics, X-rays: galaxies, line: profiles}

\abstract{ Measuring relativistic reflection is an important tool to
  study the innermost regions of the an accreting black hole
  system. In the following we present a brief review on the different
  aspects contributing to the relativistic reflection. The combined
  approach is for the first time incorporated in the new
  \texttt{relxill} model. The advantages of this more self-consistent
  approach are briefly summarized. A special focus is put on the new
  definition of the intrinsic reflection fraction in the lamp post
  geometry, which allows to draw conclusions about the primary source
  of radiation in these system. Additionally the influence of the high
  energy cutoff of the primary source on the reflection spectrum is
  motivated, revealing the remarkable capabilites of constraining
  $E_\mathrm{cut}$ by measuring relativistic reflection spectra from
  \textsl{NuSTAR}, preferably with lower energy coverage.}

\maketitle

\section{Introduction}

Reflection of X-ray radiation at the innermost regions of the
accretion disk is strongly distorted due to the strong effects of
general relativity close to the central black hole
\citep{Fabian1989}. Depending on the spin of the black hole, these
distortions will transform narrow emission lines to broad,
skew-symmetric features. This broadening is only apparent, however, if
most of the radiation originates from the innermost regions of the
accretion disk \citep{Laor1991}. Due to a combination of abundances
and fluorescent yield, the neutral Fe K$\alpha$ line at 6.4\,keV is
typically the strongest fluorescent emission line seen in AGN and GBH
spectra. First hints that the fluorescent iron line originates from
reflection of X-ray radiation at a dense accretion disk were obtained
by \citet{Barr1985a} for \cyg\ and in AGN by \citet{Nandra1989a} and
\citet{Pounds1990a}. The broadening effect of this iron line due to
the vicinity of the reflector to the black hole was already predicted
by a number of studies \citep[see, e.g.,
][]{Fabian1989,Stella1990a,Matt1991a,Matt1992a}, while the data was
still not able to constrain these models
\citep{Mushotzky1993a}. Finally, \citet{tanaka1995a} analyzed a long
\textsl{ASCA} observation of MCG$-$6-30-15 and concluded that the
skew-symmetric broad emission line present in the spectrum originates
from reflection only a few gravitational radii away from the central
black hole. Further studies have revealed that a large number of AGN
show a broadened iron line
\citep{guainazzi:06a,nandra2007a,Longinotti2008a,Patrick2011a}. With
the advent of high signal-to-noise observations (e.g., from
\textsl{XMM-Newton}, \textsl{Suzaku}, or \textsl{NuSTAR}) not only a
single broadened line could be observed, but instead the complete
reflection spectrum of the accretion disk was required to be
relativistically distorted. Those techniques lead to the detection of
relativistic reflection features in many AGN, such as, e.g.,
MCG$-$6-30-15 \citep{wilms2001a,Fabian2003b,Marinucci2014a},
1H0707$-$495 \citep{Fabian2009a,Dauser2012a}, or NGC~1365,
\citep{Risaliti2013a}, Galactic Black Holes , such as Cyg~X-1
\citep{Fabian1989,Duro2011a,Tomsick2014a}, GRS~1915$+$10
\citep{Miller2013a}, or GX~339$-$4 \citep{Miller2008a,Reis2008a}, and
neutron stars \citep[e.g., Serpens~X-1;][]{Miller2013b}.

\section{The Big Picture}
\label{sec:big-picture}

Relativstic reflection is created by a complex interplay of
effects. In the following the single contributions will be briefly
summarized. From our current understanding a primary source is
irradiating the accretion disk. The irradiating spectrum is then
reflected, imprinting the atomic features such as abundances and
ionization state of the disk into the reflected spectrum. Finally this
outgoing spectrum is relativistically distorted on its way to the
observer.

\subsection{Reflection in the Rest Frame of the Accretion Disk}
\label{sect:non_relat}

At first we will discuss the reflection itself, which
takes place in the rest frame of the accretion disk.
\begin{figure}
  \centering
  \includegraphics[width=\columnwidth]{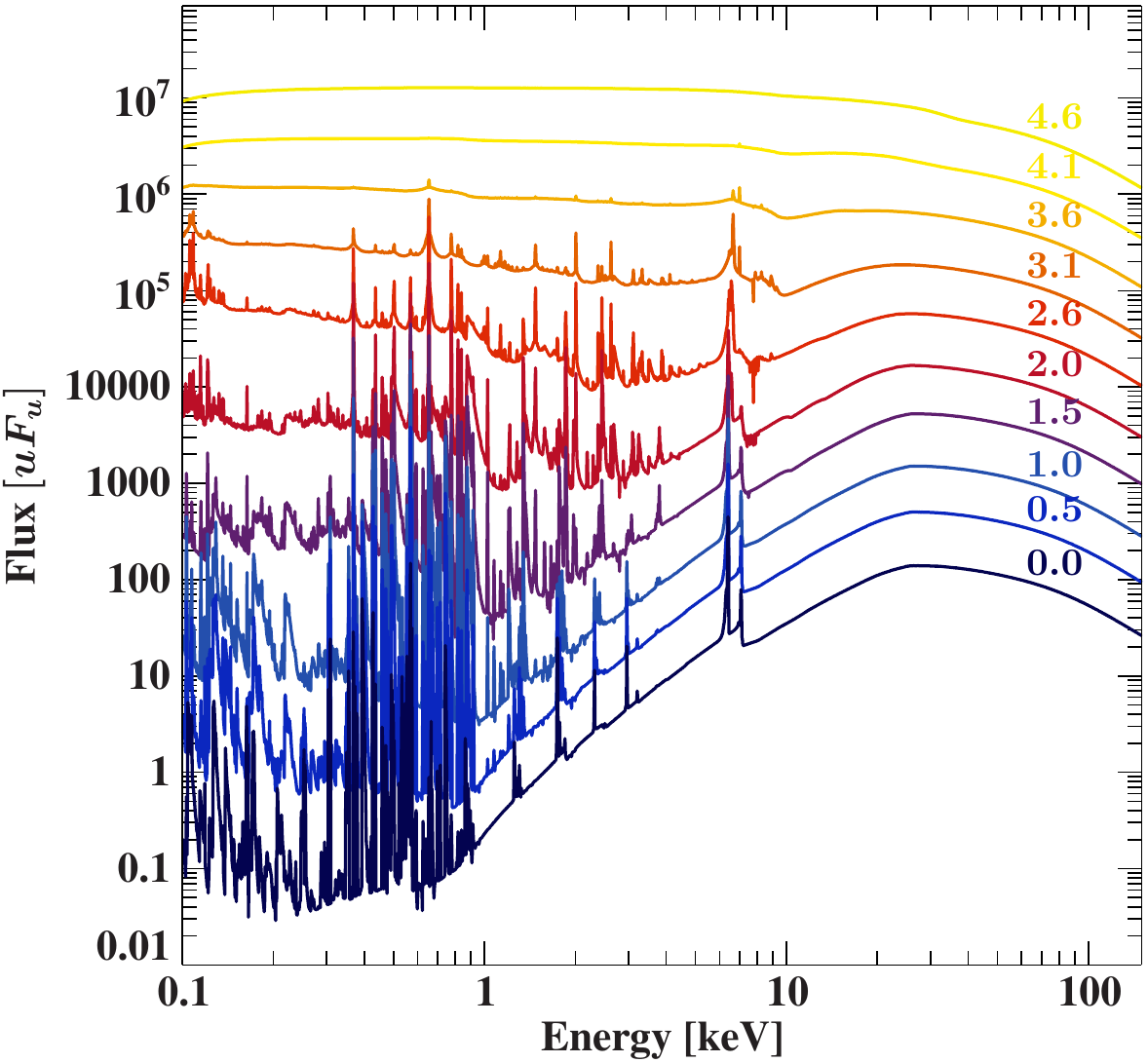}
  \caption{ Reflection spectra calculated with the \texttt{xillver}
    model \citep{Garcia2010a,Garcia2011a,Garcia2013a} for different
    values of the ionization parameter $\xi$, assuming a constant
    density of the accretion disk ($n =
    10^{-15}\,\mathrm{cm}^{-3}$). The values in the plots specify
    $\log\xi$ for each curve, from a neutral disk ($\log\xi = 0$) to a
    completely ionized disk ($\log\xi = 4.6$). Note that the spectra
    are not renormalized, but instead the flux scales linearly with
    the ionization parameter for constant density.}
  \label{fig:reflection}
\end{figure}
The model most used in the last decade for fitting relativistic
reflection by applying a relativistic blurring kernel to it, is the
\texttt{reflionx} model \citep{Ross1999a,Ross2005a,ross2007a}. More
recently, the \texttt{xillver} model
\citep{Garcia2010a,Garcia2011a,Garcia2013a} was developed. It uses a
similar approach with an extended set of atomic data. Both models are
largely in agreement. The major improvement of \texttt{xillver} comes
from the usage of the largest collection of atomic data in X-ray
astronomy, namely the \texttt{xstar}\footnote{See
  http://heasarc.gsfc.nasa.gov/xstar/xstar.html for more information.}
atomic database \citep{Bautista2001a}. One important difference
between these models concerns the assumptions made on the effect of
resonant scattering and subsequent Auger destruction. In
\texttt{reflionx}, it is assumed that this effect completely suppress
the emission of Fe K lines for second row ions (i.e., Fe~{\sc
  xvii-xxii}). In \texttt{xillver}, the branching ratios for Auger
versus fluorescence emission are explicitly calculated, and
significant line emission is still observed after the effects of Auger
decay are included.  Thus, in the range of ionization where these ions
are most predominant ($2\lesssim \log\Xi \lesssim 3$), the two models
are highly discrepant, with \texttt{reflionx} exhibiting very small Fe
K emission compared to \texttt{xillver}.

Figure~\ref{fig:reflection} shows ionized reflection spectra for
different ionizations of the accretion disk. The ionization is
parametrized by the ionization parameter $\xi$, which is defined as
the incident flux $F$ divided by the density of the disk $n$, namely $
\xi = 4 \pi F / n $. A low value implies that the disk is almost
neutral ($\xi \sim 1$). As can be seen in the figure, this leads to a
strong Fe K$\alpha$ line at $6.4$\,keV and a large forest of emission
an absorption lines at lower energies. For increasing ionization
parameter, the number and strength of these lines generally decreases,
leading to what is called a highly ionized disk for $\xi =
10^{3}$. For even larger values of ionization ($\xi = 10^{4}$), the
disk is fully ionized. This means that the disk acts almost as a
mirror and therefore the spectrum exhibits no line features
\citep[see][for a more detailed description]{Garcia2013a}. It is
notable that the most prominent feature, the iron K$\alpha$ line at
roughly 6.4\,keV, strongly changes in shape and flux for different
ionizations $\xi$.

\subsection{ Relativistic Smearing}
\label{sect:relat_smea}

\begin{figure*}
  \centering 
  \includegraphics[width=0.8\textwidth]{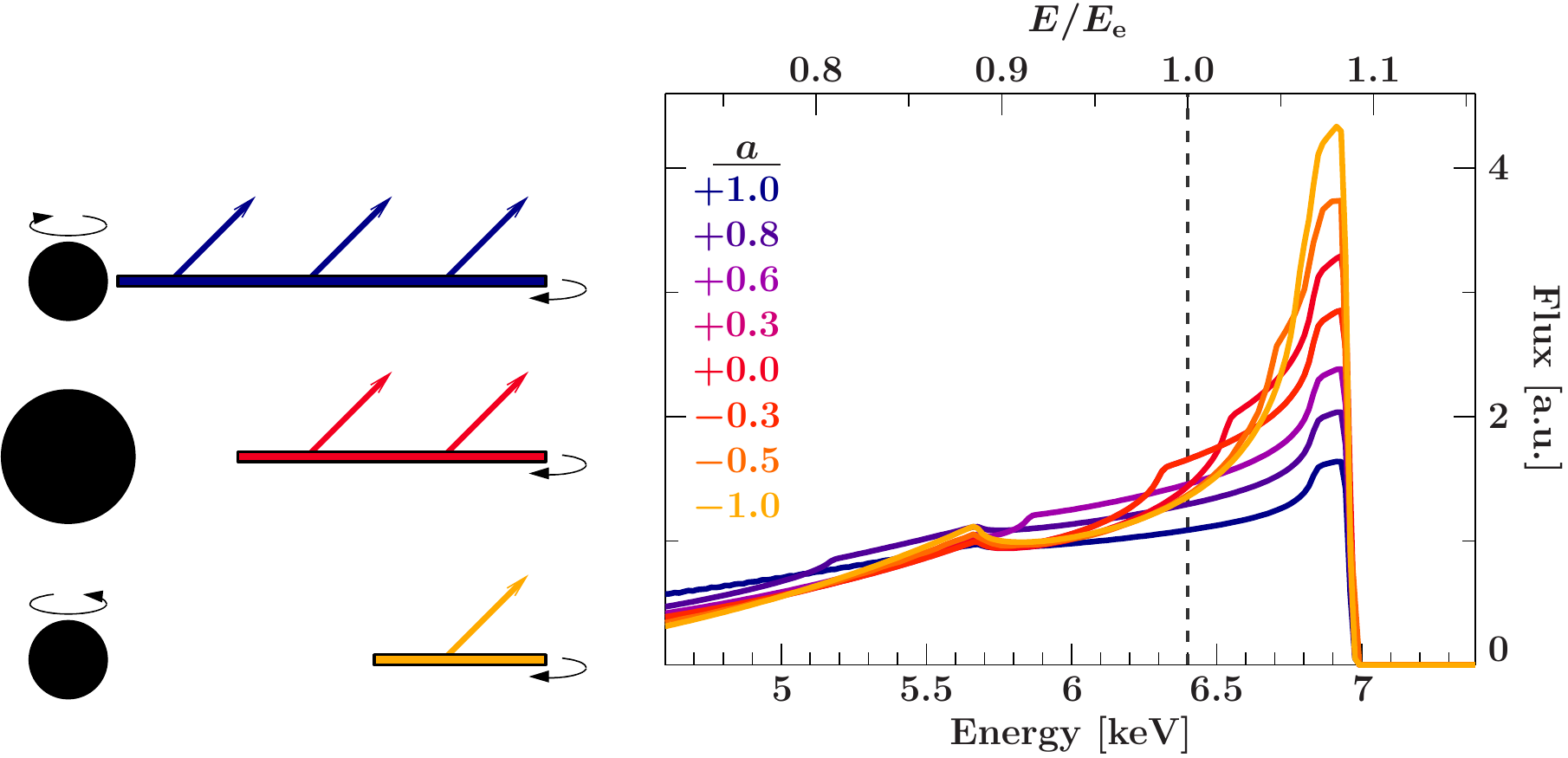}
  \caption{\emph{Left:} sketch of the black hole and the accretion
    disk for high, zero, and negative spin (from top to bottom). The
    inner edge of the disk is moving outwards for decreasing
    spin. Note that the size of the black hole changes with its
    spin. \emph{Right:} line profiles for the different values of
    spin. }
  \label{fig:line_evol}
\end{figure*}
The reflection spectrum as described in the previous section will then
get relativistically distorted on its way to the observer. Various
models have been developed in the past to describe the shape of the
relativistically smeared features (e.g., \texttt{diskline},
\texttt{laor}, \texttt{kyrline}, \texttt{kerrdisk}, or
\texttt{relline}, as published in \citealp{Fabian1989};
\citealp{Laor1991} \citealp{Dovciak2004a}; \citealp{Brenneman2006a};
\citealp{Dauser2010a}, respectively. ) All these models predict the
relativistic line profile for a narrow line emitted in the rest frame
of the accretion disk. Depending on the parameters of the system, such
as for example the spin, the irradiation of the disk, or the
inclination to the system the relativistic distortion then leads to
different line shapes. Due to the gravitational redshift and the
motion of the emitting particles in the disk, these lines are always
broadened. The actual shape of these lines strongly depends on the
parameters of the black hole system. Generally, a higher spin leads to
broader reflection features, as can be seen in
Fig.~\ref{fig:line_evol}. This is due to the fact that for larger
values of spin the accretion disk extends closer to the black hole and
therefore more highly red-shifted photons from these innermost parts
are emitted. But also other parameters such as the inclination angle
or the emissivity profile, which is the irradiating intensity
impinging on the disk, lead to significant differences in shape.

\subsection{Irradiation of the Accretion Disk}
\label{sect:irrad}

Most importantly, as shown in \cite{Dauser2013a}, the irradiating
primary source also strongly influences the broadening of the line
shape. In general terms, it determines how important the highly
relativistic inner parts of the disk contribute to the total
reflection spectrum. While a low source height leads to a strong
focusing of the photons onto the inner regions and therefore a broad
feature, a large source height irradiates a much larger part of the
disk more constantly and therefore renders the line shape to be
narrower.

In the canonical coronal model of the $\alpha$ disk
\citep{Shakura1973a}, the radial intensity is given by $I\propto
r^{-3}$. However, generally emissivities much steeper at small radii
are observed \citep[see, e.g.,
][]{wilms2001a,Fabian2004a,Dauser2012a,Miller2013a,Risaliti2013a}. The
lamp post geometry, in which the primary source is above the black
hole on the rotational axis of the black hole, automatically predicts
such steep emissivities. These considerations are strengthened by
direct measurements of the emissivity profile \citep[see
][]{Wilkins2012a}. Note that the shape of the primary source is likely
not point-like as assumed in the simple lamp post geometry, but
probably extended \citep[see][]{Wilkins2015a}. 

Independent of the origin of the primary radiation, the extent of
relativistic broadening of the reflection features already allows to
draw conclusions. Namely, the broad features can only be created by a
primary source emitting the majority of its radiation below
$10\,r_\mathrm{g}$\footnote{$r_\mathrm{g} = GM/c^2$ is the gravitational
  radius} \citep{Fabian2014a}. The fact that the primary source is
very compact and close to the black hole is also supported by
spectral-timing and reverberation studies
\citep{Kara2013a,Uttley2014a,Cackett2014a}.

In the following we present a more self-consistent approach to
modeling relativistic reflection, by focusing on its advantages
and new features.

\section{Self-consistent Modeling of Relativistic Reflection }
\label{sec:self-cons-model}

The \texttt{relxill} model\footnote{Can be downloaded at
  http://www.sternwarte.uni-erlangen.de/research/relxill/} was
implemented in order to directly combine the previously described
steps of irradiation, reflection, and relativistic smearing. This
approach has several important advantages, as has been extensively
described and analyzed in \citet{Garcia2014a}. It uses the
\texttt{xillver} model for the non-relativistic reflection (described
in Sect.~\ref{sect:relat_smea}) and the \texttt{relline} kernel for
the relativistic blurring.

The most important change is that the new model allows for the first
time to take the angular directionality in a relativistic reflection
spectrum into account. As is described in detail in
\citet{Garcia2014a} this means that instead of taking an angle
averaged spectrum for the complete accretion disk, now each point of
the disk gets the proper reflection spectrum for the emission angle as
seen by the distant observer. Note that due to strong gravitational
light-bending effects this emission angle often strongly differs from
the inclination angle, which necessitates the chosen approach.

As also presented in \citet{Garcia2014a}, the direct consequence of
the change in spectrum is that important system parameters like the
spin and the inclination angle are better constrained by the
\texttt{relxill} model. The actual values of those parameters is
mainly consistent with the previous, angle-averaged approach. Solely
we could show that the iron abundance might have been over-estimated
by up to a factor of two. 

Besides the direct consequences of the self-consistent model by taking
the angular directionality into account, the combined approach of the
\texttt{relxill} model does have additional important advantages. In
the following we will briefly focus on the reflection fraction as a
new, intrinsic parameter (Sect.~\ref{sec:reflection-fraction}) and
the possibility to constrain coronal parameters by exploiting the strong
influence of the high energy cutoff on the reflection spectrum
(Sect.~\ref{sec:high-energy-cutoff}).

\subsection{The Reflection Fraction}
\label{sec:reflection-fraction}

The \texttt{relxill} model allows for the possibility to predict the
direct and reflected spectrum simultaneously. While strictly speaking
no improvement in the spectral model itself, this combination has
important consequences by allowing to fit directly for the reflection
fraction $R_\mathrm{f}$. This reflection fraction, which characterizes
the strength of the reflection, can be used as a measure for the
accretion geometry. As shown in \citet{Dauser2014a}, it also allows
for stricter constraints on the spin of the black hole, as high values
of the reflection fraction are only possible for a rapidly rotating
black hole.

\subsubsection{Definition}
Formally, the reflection fraction is defined in terms of the fraction
of emitted photons of the primary source which will hit the accretion
disk ($N_\mathrm{AD}$)compared to the fraction of photons escaping to
infinity ($N_\infty$):
\begin{equation} \label{eq:1}
  R_\mathrm{f} = \frac{N_\mathrm{AD}}{N_\infty}
\end{equation}
In order to consistently calculate these fractions, a certain geometry
has to be assumed. In the case of the lamp post geometry, these
fraction have been calculated in a ray tracing simulation as described
in \citet{Dauser2014a}.

The important fact about this definition of the reflection fraction is
that it is not simply the ratio of observed fluxes, but it is
intrinsic to the black hole system. Therefore we can directly measure
the fraction of photons hitting the accretion disk, making it possible
to put constraints on the geometry of the system. Such an intrinsic
definition is currently only available for the lamp post geometry
(i.e., the \texttt{relxilllp} model). As motivated in
Sect.~\ref{sect:irrad}, however, the lamp post geometry seems to be a
good description for the primary source in the observed sources.

The reflection fraction of the normal \texttt{relxill} model, which
does not assume a certain geometry, is defined by simply taking the
ratio of the reflected flux to the direct flux in the 20--40\,keV
band. This energy range, coinciding with the Compton hump, is chosen
as its flux is largely independent of the intrinsic parameters of the
reflection such as the ionization or the abundance (see
Fig.~\ref{fig:reflection}).\footnote{Note that this definition also
  differs from the one of the \texttt{pexrav} model, which uses the
  total reflected flux to define the reflection fraction. While being
  a good description for distant reflection, it is therefore not
  applicable for relativistic reflection.} This approach, however,
still does not take any relativistic boosting effects into account and
therefore this value of reflection fraction might largely differ form
the intrinsic definition in some cases.

\subsubsection{Overall Flux Normalization of \texttt{relxilllp}}
In order to predict the flux starting from the intrinsic reflection
fraction, we need to ensure a proper normalization throughout the
ray-tracing code. As flux is not directly conserved in general
relativity (see Liouville's theorem), additional effort has to be
taken in order to ensure this. The different steps will be briefly
summarized in the following.

Such a definition of the normalization then allows to directly compare
the flux for different combinations of parameters of the
\texttt{relxilllp} model (e.g., for different source
heights).\footnote{Feature is included since the \texttt{relxill}
  version v0.3a.}

\paragraph{Irradiation}
Starting with the intensities emitted at the primary source
($N_\mathrm{AD}$ and $N_\infty$, see Eq.~\ref{eq:1}), we calculate the
intensity as seen at infinity ($I_\infty$) and at the accretion disk
($I_\mathrm{AD}$). In the non-relativistic limit we would expect an
equal number of photons emitted towards the disk and towards infinity
(i.e., $N_\mathrm{AD}$ = $N_\infty$).

Using the momentum and 4-velocity presented in \citet{Dauser2013a},
the direct radiation observed at infinity from a source at height $h$
is then given by
\begin{equation} \label{eq:2}
  I_\infty = \left(\frac{1}{u^t_\mathrm{h}}\right)^\Gamma 
  N_\infty =  \left( \sqrt{ 1 - \frac{2h}{h^2 + a^2}  }  \right)^\Gamma   N_\infty
\end{equation}
Similarly, as calculated in \citet{Dauser2013a}, the incident
intensity on the accretion disk is given by
\begin{equation}
I_\mathrm{AD} = \frac{\sin \de \gi^{\pli} }{A(r,\dr)\gaphi}  N_\mathrm{AD}
\end{equation}
with $\delta$ being the emission angle at the primary source, $A$ the
effective area of a ring $\Delta r$ as seen by the incident photons,
$\gaphi$ the Doppler factor, and
\begin{equation}\label{eq:energy-shift}
  \gi = 
  \frac{\left(\ri\sqrt{\ri}+a\right)\sqrt{h^2-2h+a^2}}
{\sqrt{\ri}\sqrt{\ri^2-3\ri+2a\sqrt{\ri}}\sqrt{h^2+a^2}} \quad .
\end{equation}

\begin{figure}
  \centering
  \includegraphics[width=\columnwidth]{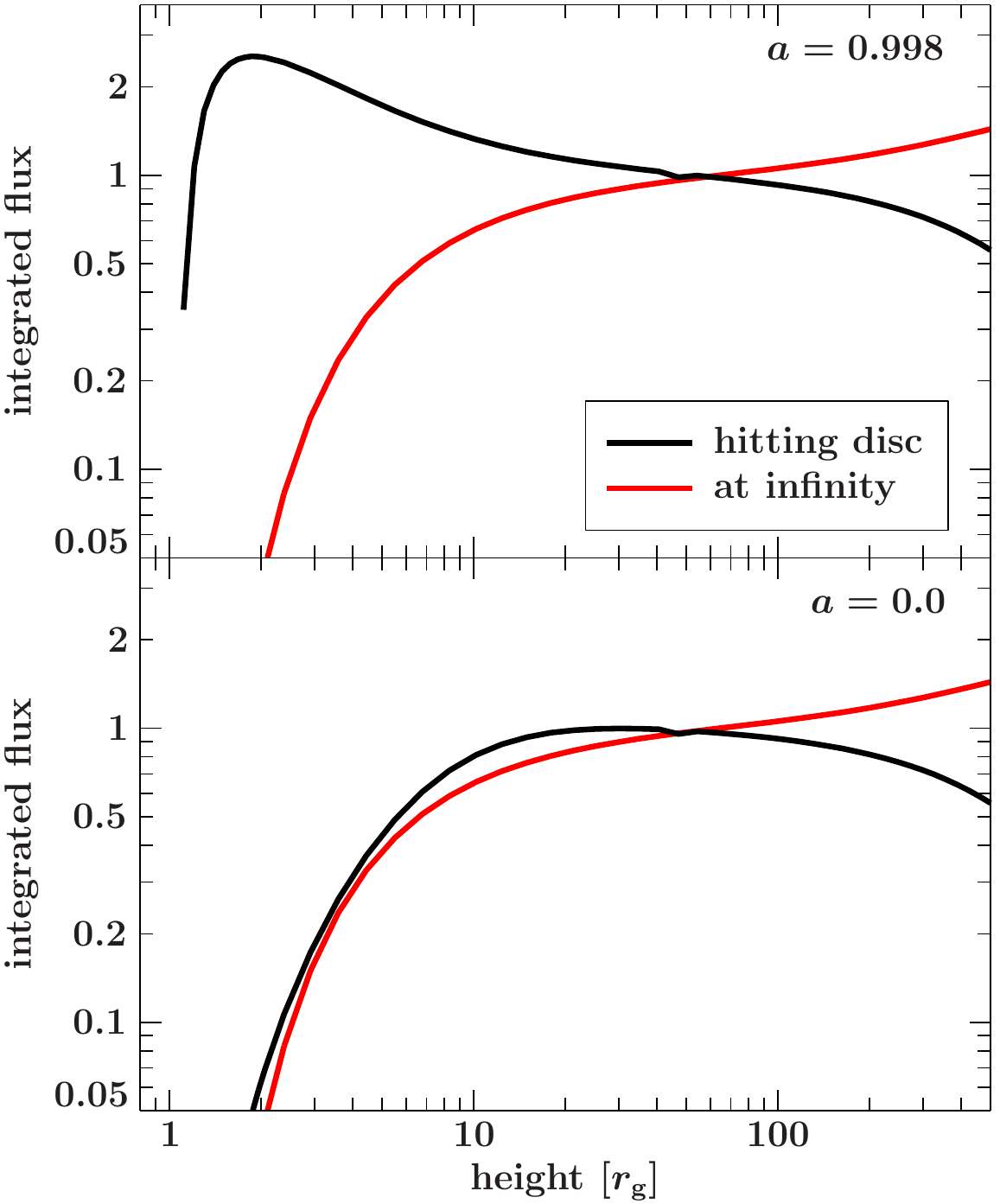}
  \caption{Plotted is the flux hitting the disk (black) and the flux
    at infinity (red). The flux at infinity is reduced at low heights
    due to gravitational red-shift.  One can also nicely see that for
    low spin and low source height basically not much reflection and
    even less direct radiation is expected.  This plot does not yet
    include any effects from reflection at the disk and from the disk
    to the observer (boosting and inclination).  Note that the
    behavior at large heights is exactly what one would expect from
    an accretion disk with an outer radius of 1000\,$\rg$.  }
  \label{fig:frac}
\end{figure}
Using these information and the fraction calculated from the incident
intensity at the accretion disk and at infinity for the direct
radiation are plotted in Fig.~\ref{fig:frac}. Note that the
normalization condition is full-filled although the curves do not
converge towards 1. First of all the accretion disk in this simulation
is truncated at an outer radius of 1000\,\rg\ and therefore for larger
heights the direct flux as seen at infinity increases simply due to
geometrical reasons. Moreover, it is also evident that even for large
heights the total integrated flux does not add up to a value of 2,
which would be expected in flat space time. The reason for this is
that the energy boosting of the photons hitting the inner parts of the
accretion disk increase the overall flux even for large source heights
due to boosting from the energy shift\footnote{Photons emitted from
  large distances are seen strongly blue-shifted at the inner parts of
  the accretion disk, which also leads to a strong boost in
  flux}. Note that the strength of this boosting strongly depends on
the steepness of the incident spectrum \citep[see][]{Dauser2013a}.

\paragraph{Reflection}
The flux of the incident radiation ($I_\mathrm{AD}$) is now used to
normalized the intrinsic \texttt{xillver} spectrum.  For this purpose
we need the normalization of the incident spectrum, which was used to
calculate the \texttt{xillver} table. As described in, e.g.,
\citet{Garcia2013a} it is defined as a cutoff power law with power law
index $\Gamma$ and an exponential high energy at an energy of
$E_\mathrm{cut}$. The total flux of the input radiation is set to be
\begin{equation}
 \int_{E_\mathrm{lo}}^{E_\mathrm{hi}} I_\mathrm{incident} = \frac{\xi
   n}{4 \pi} \quad, 
\end{equation}
where $E_\mathrm{lo}=0.1\,$keV, $E_\mathrm{hi}=1000\,$keV, the density
$n=10^{15}\,\mathrm{cm}^{-2}$, and $\xi$ the ionization parameter.

\paragraph{Relativistic Smearing}
As a last step, the reflection spectrum in the rest frame of the disk,
which is already normalized regarding the incident flux, is
relativistically smeared on its way to the observer. Intrinsically,
this is achieved by the \texttt{relconv} kernel. The emitted flux
integrated over the complete area of the accretion disk is
automatically conserved for the non-relativstic limit by using the
chosen approach with the Cunningham transfer function \citep[see,
][]{Cunningham1975,Speith1995,Dauser2010a}. Relativistic effects alter
the observed flux in the expected way, by Doppler boosting and general
relativistic red-shift. In addition to these effects the inclination
to the accretion disk also plays an important role here. The intensity
due to the projection on the sky follows roughly
$\cos\theta_\mathrm{obs}$. Relativistic light-bending slightly changes
the picture. Especially the flux behind the black hole (as seen from
the observer) is enhanced for large inclination angles.

The \texttt{relxilllp} model now returns this reflected flux together
with the direct radiation from the primary source, which is given by
Eq.~\ref{eq:2}. 

\subsection{The High Energy Cutoff}
\label{sec:high-energy-cutoff}

Besides the reflection fraction, measuring the high energy cutoff
$E_\mathrm{cut}$ of the primary power law also allows to constrain the
accretion geometry. Its value is strongly connected to the temperature
of the corona, which also allows to draw conclusions about its size
and origin. Especially since the launch of \textsl{NuSTAR}
\citep{Harrison2013a}, several good measurements of $E_\mathrm{cut}$
have been obtained \citep[see,
  e.g.,][]{Brenneman2014a,Ballantyne2014a}. However, as the
\textsl{NuSTAR} energy range is 3--79\,keV, constraining the high
energy cutoff using the direct measurement of the continuum radiation
only works up to 200\,keV. As has been shown in \citet{Garcia2015a},
remarkably the relativistic reflection is easily capable in
constraining the cutoff energy up to 1000\,keV. This is due to the
fact that the cutoff of the primary spectrum has a distinct effect on
the reflection spectrum around the Compton hump (20--40\,keV) and also
at energies below 10\,keV. 

\citet{Garcia2015a} show in explicit simulations the capabilities to
constrain the cutoff with the reflection method. Additionally it is
shown that only applying the cutoff on top of the reflection spectrum
yields poor and systematically biased results. Therefore using the
cutoff as built in by \texttt{xillver} and \texttt{relxill} is
essential for accurate results. A main conclusion of the aforementioned
publication is that while only \texttt{NuSTAR} data can only set
constraints when $E_\mathrm{cut}<400\,$keV. Combined datasets, e.g.,
by adding \textsl{Suzaku} to the \textsl{NuSTAR} data, allows to
constrain the energy of the cutoff of 1\,mCrab sources far beyond the
energy range of the detectors. The highly increased sensitivity below
10\,keV is responsible for this effect and clearly demonstrates the
impact of the cutoff energy on the complete reflection spectrum.

Using now the \texttt{relxill} setup, additional information of the
source in terms of height/compactness and reflection strength can be
obtained from the same data. A systematic study of these parameters
will allow us to draw firmer conclusions on the geometry of the source
and test coronal models. The main challenge for those models is to
predict the huge amount of flux emitted at the compact primary
sources, which are found from timing (reverberation) and spectral
(relativistic reflection) measurements.

\section{Summary and Conclusions}
\label{sec:summary-conclusions}

Relativistic reflection is a complicated interplay between
irradiation, the reflection in the rest frame of the disk, and the
smearing of the reflected spectrum on its way to the observer. Besides
the spectrum also intensities strongly change. Especially strong
reflection features are created by compact primary sources, which
irradiate an accretion disk around a rapidly rotating black hole. The
newest version of the \texttt{relxill} model does incorporate all of
these effects. By choosing the lamp post geometry it is possible to
directly fit intrinsically the irradiating source. The characterizing
parameter is the reflection faction, which is defined by the ration
between photons emitted towards the disk and towards the observer.

Additionally we motivated that the high energy cutoff of the direct
radiation can be measured indirectly by fitting relativistic
reflection. It was shown that for values far beyond the observed
energy band strong constraints can be put on $E_\mathrm{cut}$
\citep{Garcia2015a}. Especially good results are obtained when
combining high energy data from \textsl{NuSTAR} with detectors at
lower energies, such as \textsl{Suzaku}.

Having now tools at hand for directly analyzing the geometry of the
accreting systems, a systematic study of these objects and their
properties are the next important step for understanding the extremes
of the accretion at the smallest scales.

\acknowledgements This research has made use of ISIS functions
provided by ECAP/Remeis observatory and MIT
(http://www.sternwarte.uni-erlangen.de/isis/). We gratefully thank
John E. Davis for developing the \texttt{SLXfig} module, which was used
to prepare all figures in this publication.


\end{document}